\documentclass[12pt]{iopart}
\usepackage{graphicx}
\begin{document}
\title{The Quadratic Spin Squeezing Operators}
\author{P. K. Pathak$^1$, R. N. Deb$^2$, N. Nayak$^3$, and B. Dutta-Roy$^3$}
\address {$^1$Department of Physics, Chonnam National University,
Gwangju - 500757, South Korea\\
$^2$ Physics Department, Krishnagar Government College, Krishnagar,
West Bengal, India\\
$^3$S. N. Bose National Centre for Basic Sciences, Block-JD, Sector-3,
Salt Lake City, Kolkata-700098, India}
\begin{abstract}
We discuss generic spin squeezing operators (quadratic in angular momentum operators)
capable of squeezing out quantum mechanical noise from a system of two-level atoms
(spins) in a coherent state. Such systems have been considered by Kitagawa and Ueda
(Refs. 2) in this context and a Hamiltonian of this nature governs the Lipkin model
(Ref. 14) which is relevant to nuclear physics.
\end{abstract}
\submitto{\JPA}
\maketitle
\section{Introduction}
We consider a system of $N$ two-level atoms (TLA) and represent them
by the algebra of spin one-half particles with total angular momentum
$\mathbf J$ with $j=N/2$. Hence, we refer to the system as a system of
$N$ spins or TLA. We start with the system in a coherent
state [1]
\begin{eqnarray}
\vert j,\xi\rangle & = &
e^{\xi J_{-}-\xi^{*}J_{+}}\vert j,m=+j\rangle \nonumber \\
& = & \frac{1}{(1+\vert\xi\vert^2)^j}\sum_{n=0}^{2j}\sqrt{^{2j}C_n}\xi^n\vert
j,m=j-n\rangle
\end{eqnarray}
where $\xi$ is a complex number given by
$$\xi =\tan (\theta /2)e^{i\phi}. $$
$\theta =0$ represents the state with all the atoms in the upper of the two states
(spin-up). The coherent state satisfies the equality sign in the uncertainty relation
for the components of `angular momentum'
in a plane normal to the mean-spin vector $\langle\mathbf J\rangle$ with equal amount
of noise in its two quadratures. The mean-spin vector has the magnitude
\begin{equation}
\vert\langle{\mathbf{J}}\rangle\vert =
\sqrt{\langle J_x\rangle^2 +\langle J_y\rangle^2 +\langle J_z\rangle^2}.
\end{equation}
In other words, in a frame in which $\langle\mathbf J\rangle$ is along an axis
($z^\prime$ say), the minimum uncertainty condition is satisfied by the state described
by Eq. (1), viz.,
\begin{equation}
\Delta J_{x^\prime} = \Delta J_{y^\prime} = \sqrt{\vert\langle{\mathbf J}\rangle\vert/2}
\end{equation}
where $\Delta J_{x^\prime,y^\prime}=\sqrt{\langle J_{x^\prime,y^\prime}^2\rangle -
\langle J_{x^\prime,y^\prime}\rangle^2}$ are the variances in the two quadratures.
If in such a frame the condition
\begin{equation}
\Delta J_{x'}\sqrt{2/\vert\langle{\mathbf J}\rangle\vert} ~~or~~
\Delta J_{y'}\sqrt{2/\vert\langle{\mathbf J}\rangle\vert} < 1
\end{equation}
is satisfied in some state of the system, then we shall say that the atomic system is
spin squeezed. \\

Model Hamiltonians through which spin squeezed states could be realised were proposed
by Kitagawa and Ueda [2] and others [3]. Since then, it has been vigorously studied
in various atom-field systems [4-12]. When a TLA interacts with a
squeezed radiation field, squeezing can be transferred from the field to the atomic
system [4]. Self squeezing of the spin system is possible if it is adequately nonlinear
in its interaction with the radiation field [5]. The well-known Tavis-Cummings model
also exhibits spin squeezing [6]. Its importance has been enhanced further due to its
potential application in the area of quantum information [7-10].\\

We have been studying various quantum mechanical operators, rather nonlinear in
spin operators, capable of producing spin squeezed states when applied to a TLA.
In Ref. [11], spin squeezing properties of a pseudo-Hermitian operator
\begin{equation}
\Lambda =\left(e^{\epsilon}J_{+} + e^{-\epsilon}J_{-}\right)/2
\end{equation}
with real $\epsilon$ have been studied. The operator $\Lambda$ has the features of
pseudo-Hermicity as $\Lambda^\dagger\ne \Lambda$ but yet
$\Lambda^\dagger=\eta\Lambda\eta^{-1}$ where $\eta$ is linear
invertible Hermitian operator. Further, $\eta$ can be written as $\eta =O^\dagger O$
where $O=O^\dagger = \exp (-\epsilon J_z)$ and, so, the eigenstates of $\Lambda$
has real eigenvalues.
This operator is of interest in the study of a TLA
interacting with a squeezed vacuum [4]. Studies in [2, 12] reveal
that with a Hamiltonian quadratic in the spin operator $J_z$, that is,
\begin{equation}
\hat{H_1}=-\Gamma J_{z}^2
\end{equation}
(where $\Gamma$ is a constant characteristics of the system) the corresponding evolution
operator $\exp(-i\Gamma t J_{z}^2)$ when
applied to a spin coherent state in Eq. (1) produces spin squeezing
in the TLA. This operator can describe the interaction of a TLA with
radiation field in a high-Q dispersive cavity [13]. This operator is
also a special case of the so-called
Lipkin-Meshkov-Glick (LMG) Hamiltonian for a many-body fermionic system [14].\\

In this paper we identify the quadratic atomic (spin) operator which can introduce a squeeze from an analogy between the spin and bosonic systems.
We do so in Sec. II and study its squeezing properties in the following
sections. In Sec. III, we give analytical results for a two-atom system
which provide some insight into the properties of the proposed
operator. The numerical results for $N>2$ are discussed in the Sec.
IV. We conclude the paper in Sec. V with a comment on possible realization
of such operators in some physical
systems.\\

\section {The operator}
We begin with the derivation of a displacement operator for the spin coherent state
in Eq. (1).
This can be achieved if we cast the state $\vert j,\xi\rangle$ in the Schwinger
representation [15]. The spin operators in the Schwinger representation are constructed by
defining bosonic annihilation operators for two modes $a_i (i=+,-)$ such that
$[a_i ,a_{j}^\dagger]=\delta_{ij}$ and $[a_i ,a_{j}]=0=[a_{i}^\dagger ,a_{j}^\dagger]$.
They take the forms $J_+=a_{+}^\dagger a_-$, $J_z =[a_{+}^\dagger a_{+}
-a_{-}^\dagger a_{-}]/2$, etc.
The spin coherent state $\vert j,\xi\rangle$ takes the form
\begin{eqnarray}
\vert j,\xi\rangle &=& \frac{1}{(1+\vert\xi\vert^2)^j \sqrt{2j!}}
         \sum_{n=0}^{2j}{^{2j}C_n}\xi^n(a_{+}^\dagger)^n(a_{-}^\dagger)^{2j-n}
                                \vert0_+ ,0_-\rangle   \nonumber \\
                    &=& D_{+,-}\vert0_+ ,0_-\rangle
\end{eqnarray}
where $\vert0_+ ,0_-\rangle$ are vacuum (fictitious) states for $+$ and $-$ modes.
Comparing this form of the spin coherent state with the Glauber-Sudarshan coherent state for bosonic
particles [16]
\begin{equation}
\vert\alpha\rangle =D(\alpha)\vert 0\rangle
\end{equation}
where $D(\alpha)$ is the well-known displacement operator
\begin{eqnarray}
D(\alpha) &=& \exp(\alpha a^\dagger -\alpha^\ast a) \nonumber \\
          &=& e^{-\vert\alpha\vert^2 /2}\sum_{n=0}^\infty\frac{\alpha^n (a^\dagger)^n}{n!}
\end{eqnarray}
it may be said that $D_{+,-}$  via the Schwinger construction is a `displacement' operator for
spin systems. \\


The above analogy between Glauber-Sudarshan coherent states and the spin coherent states suggest
that one may attemp to obtain spin squeezed states in a manner similar to that employed for the
bosonic case. Thus a quadratic form of the generators $a$ and $a^\dagger$ of the Heisenberg-Weyl
algebra was used to set up the operators which when acting on the Glauber-Sudarshan coherent
states resulted in the squeezed states, viz, taking a cue from [17],
\begin{equation}
S_{bosonic}=\exp[\frac{1}{2}(\chi^\ast a^2 -\chi a^{\dagger 2})] .
\end{equation}
Likewise, for the atomic or (pseudo)-spin coherent states it is tempting to consider exponentiation
of the quadratic form constructed out of the generator $J_x$, $J_y$, and $J_z$ of the rotation
group. Consider the most general such form, namely, $J_{k}J_{l}$ where $l.k$ run from $x,y,z$.
Note that $J_l$ transforms like a vector operator (in the 3-dimentional representation)
under rotation and accordingly $J_lJ_k$ is a tensor of the second rank. However, this may be
reduced in the following manner
\begin{eqnarray}
J_{l}J_{k} &=& \big[\frac{1}{2}(J_{l}J_{k}+J_{k}J_{l})-\frac{1}{3}\delta_{lk}J^2\big] \nonumber \\
           &+&\big[(\frac{1}{2}(J_{l}J_{k}-J_{k}J_{l})\big]+\big[\frac{1}{3}\delta_{lk}J^2\big]
\nonumber
\end{eqnarray}
The first term in square brackets being the components of a traceless symmetric second rank tensor
(with five independent components), the second being the components of the antisymmetric
second rank tensor which by virtue of the commutation relation $[J_k,J_l]=i\epsilon_{lks}J_s$ is
expressed in terms of the vector representation and the last term in square backets is a
scalar $J^2$. This is tentamount to the reduction of the direct product of two vectors
$$ 3\otimes 3 =5 \oplus 3 \oplus 1.$$
Thus the exponentiation in $\exp{i\eta J_kJ_l}$ being quadratic in the generation of the rotation
group can be decomposed into the sum of three operators with the trace
$\sum J_kJ_l\delta_{kl}=J^2$ yielding a mere overall phase, the antisymmetric form in the exponent
just a rotation, and the quadratic form in a exponent yielding the non-trivial squeeze. Since
the different components of the quadrupole tensor are related via rotations we can in effect
express the squeezing operator in the case of the spin system choosing, for example,
the fiducial quadratic forms
\begin{eqnarray}
S_{spin}&=&\exp[\eta (J_{x}J_{y}+J_{y}J_{x})]    \nonumber \\
S^{\prime}_{spin}&=&\exp[\eta J_{z}^2]
\end{eqnarray}
Thus $S_{spin}$ can represent time evolution of a state vector under the action of a Hamiltonian
\begin{equation}
H=\zeta [J_{x}J_{y}+J_{y}J_{x}]
\end{equation}
where $\eta =-i\zeta t$. \\

Indeed, $H$ is connected to other quadratic combinations of spin
operators, viz, $[J_{y}J_{z}+J_{z}J_{y}]$ and $[J_{x}J_{z}+J_{z}J_{x}]$ by mere rotations.
Similarly, $\hat{H_1}$ in Eq. 6, ${J_x}^2$ and ${J_y}^2$ are related to one another by simple
rotations. Thus the operators in Eq. (11) can be taken to be the representatives of
interaction Hamiltonians having quadratic forms of spin operators. Kitagawa and Ueda [2] has
shown that the operator $S^{\prime}_{spin}$ in Eq. (11) produces spin squeezing resulting from a
single-axis twisting and the operator $S_{spin}$ in Eq. (11) or the Hamiltonian in Eq. (12)
can give rise to spin squeezing resulting from two-axis counter twisting.  We, however,
present a detail account of this operator and show that a large number of atoms (spins) can
be squeezed. We also point out the possibilities of its physical realization.\\

In Ref. 12 we have already studied the spin squeezing dynamics of $S^{\prime}_{spin}$.
What follows is a discussion on spin squeezing properties of the operator $S_{spin}$.\\

We can write $J_+$ and $J_-$ in the Schwinger representation and
proceed to examine the spin squeezing properties when operated upon a spin coherent state.
This may produce a spin squeezed state
$$ \vert sss\rangle =S_{spin}D_{+,-}\vert0_+ ,0_-\rangle $$
in the lines of Yuen's representation for a bosonic squeezed state [17]. However, this method
gets very complicated and, so, we investigate the squeezing properties of $S_{spin}$ as
follows. \\

We assume that the operator acts on a system of atoms prepared initially in a coherent state
$\vert\theta ,\phi\rangle\equiv\vert j,\xi\rangle$. This can be achieved by sending the atoms
through a cavity whose single-mode is maintained by the radiation from a laser. The polar
angle $\theta$ is decided by the interaction time of the atoms with the cavity field. We
first consider the action of $S_{spin}$ on a two-atom coherent state as it gives analytical
results. With the insight gained from these results, we go ahead with the numerical analysis
of squeezing properties of  $S_{spin}$ acting on a system having more than two atom.\\

\section {A two-atom system}
In addition to this being the first step of our general study, the bipartite system is of
special interest in the study of quantum entanglement [7-10]. \\

The initial state of the system, the coherent state, is in this case given by
\begin{equation}
\vert\xi ,\theta ,\phi\rangle =\frac{1}{(1+\vert\xi\vert^2)}\Big[\vert 1,+1\rangle +
                                \sqrt{2}\xi\vert 1,0\rangle +\xi^2\vert 1,-1\rangle \Big]
\end{equation}
which is a linear superposition of the three Wigner state with
$m=0,\pm 1$. The action of $S_{spin}$ on the $\vert\xi ,\theta
,\phi\rangle$ changes the amplitudes of the Wigner states giving
\begin{eqnarray}
\vert sss\rangle &=& S_{spin}\vert\xi ,\theta ,\phi\rangle        \nonumber \\
                 &=& C_1 \vert 1,+1\rangle +C_2\vert 1,0\rangle +C_3\vert 1,-1\rangle
\end{eqnarray}
where
$$C_1 =\frac{1}{(1+\vert\xi\vert^2)}\Big[\cos (2\vert\eta\vert)+\sqrt{\eta/\eta^\ast}
                                          \xi^2\sin (2\vert\eta\vert)\Big],$$
$$C_2=\frac{\sqrt{2}\xi}{(1+\vert\xi\vert^2)},$$
and
$$C_3 =\frac{1}{(1+\vert\xi\vert^2)}\Big[\xi^2\cos (2\vert\eta\vert)-\sqrt{\eta^\ast/\eta}
                                          \sin (2\vert\eta\vert)\Big].$$

It has been shown in [18] that this state radiates squeezed light [17], thus,
indicating strongly that the collective two-atom state is spin squeezed [5].
However, its spin squeezing properties, as it is known by now [2-12] and
defined in Eq.(4), have not been discussed there. We present them below.

It is straightforward to calculate the moments and correlation
functions of the spin operators. We find that all the three
correlation functions, namely $\langle J_xJ_y + J_yJ_x\rangle$,
$\langle J_xJ_z +J_zJ_x\rangle$ and $\langle J_yJ_z
+J_zJ_y\rangle$ are nonzero and for real $\xi$ and $\eta$ one of
them, $\langle J_xJ_z +J_zJ_x\rangle$, survives. This indicates the
squeezing capabilities of $S_{spin}$ for all possible values of
$\xi$ and $\eta$. This is due to the fact that nonzero values of
correlation function (at least one) is a necessity for spin
squeezing to take place [2,11]. The variances in the two
quadratures, namely $x^\prime$ and $y^\prime$, in the rotated frame
as defined in Eqs. (2-4) are given  by
\begin{eqnarray}
(\Delta J_{x^\prime})^2 &=& \frac{1}{\vert\langle{\mathbf J}\rangle\vert^2} \bigg[
                \frac{1}{2(1+\xi^2)^4}\bigg\{
                                             (\xi^8-2\xi^4+1)\cos^2(4\eta)-     \nonumber \\
          & &2\xi^2(\xi^4-1)\sin(8\eta)+4\xi^4\sin^2(4\eta)
                                        \bigg\}+       \nonumber \\
             & &\frac{1}{(1+\xi^2)^6}\bigg\{(3\xi^{10}-10\xi^6+3\xi^2)\cos^3(4\eta)- \nonumber \\
             & & 3(\xi^{10}-2\xi^6+\xi^2)\cos^2(4\eta)-2(\xi^{10}-4\xi^8- \nonumber \\
             & & 4\xi^6-4\xi^4+\xi^2)\cos(4\eta)-\frac{1}{2}(\xi^{12}-15\xi^8+ \nonumber \\
             & &15\xi^4-1)\sin^3(4\eta)-12\xi^6\sin^2(4\eta)+\frac{1}{2}(\xi^{12}+ \nonumber \\
             & &8\xi^{10}-11\xi^8+11\xi^4-8\xi^2-1)\sin(4\eta)+6\xi^4\times \nonumber \\
             & &(\xi^4-1)\sin(8\eta)+4\xi^{10}+8\xi^6+4\xi^2 \bigg\}\bigg]
\end{eqnarray}
and
\begin{eqnarray}
(\Delta J_{y^\prime})^2 &=& \frac{1}{2}\bigg[1+\frac{1}{(1+\xi^2)^2}\bigg\{
                               2\xi^2(1-\cos(4\eta))+      \nonumber \\
          &&(1-\xi^4)\sin(4\eta)\bigg\}\bigg] .
\end{eqnarray}
In Eq. (15), the presence of $\vert\langle{\mathbf J}\rangle\vert^2$ in the denominator is due to the
rotation required to bring $\langle{\mathbf J}\rangle$ along the required direction, here, $z^\prime$
axis. The mean-spin vector is given by
\begin{eqnarray}
\vert\langle{\mathbf J}\rangle\vert^2 &=& \frac{1}{(1+\xi^2)^4}\bigg[4\xi^2\bigg\{
                      1+\xi^4+2\xi^2\cos(4\eta)-(1-\xi^4)\times \nonumber \\
                 && \sin(4\eta)\bigg\}+ (1-\xi^4)^2\cos^2(4\eta)+4\xi^4\sin^2(4\eta)- \nonumber \\
            &&2\xi^2(\xi^4-1)\sin(8\eta)\bigg]
\end{eqnarray}
It is easy to note that for $\eta=0$ the results reduce to that for a spin coherent state, that is,
$\vert\langle{\mathbf J}\rangle\vert =1$ and $\Delta J_{x^\prime}=\Delta J_{y^\prime}=1/\sqrt{2}$ in
accordance with Eqs. (2) and (3) for a two-atom system. Further, we notice in Eqs. (15) and (16) that
the arguments of all the circular functions appearing there are in the form of $4\eta$ and $8\eta$.
Hence, the variances return to their initial condition , that is a coherent state, for $\eta=n\pi/2$
where $n$ is an integer. The spin squeezing parameters $S_{x^\prime}$ and $S_{y^\prime}$ as defined in
Eq. (4)
\begin{equation}
S_{x^\prime ,y^\prime}=\Delta J_{x^\prime ,y^\prime}\sqrt{2/\vert\langle{\mathbf J}\rangle\vert}
\end{equation}
has interesting relations for $\theta=0$ and $\pi$:
\begin{equation}
S_{x^\prime}\vert_{\theta=0}=S_{y^\prime}\vert_{\theta=\pi} =\sqrt{\frac{1+\sin(4\eta)}{\cos(4\eta)}}
\end{equation}
and
\begin{equation}
S_{x^\prime}\vert_{\theta=\pi}=S_{y^\prime}\vert_{\theta=0} =\sqrt{\frac{1-\sin(4\eta)}{\cos(4\eta)}}.
\end{equation}
Because of the above observations it is sufficient to study the variances for $0<\eta\le\pi/2$.
We display them in Fig. 1 and notice perfect squeezing of the two-spin system.
\vskip 0.4cm
\begin{figure}[h]
        \begin{center}
        \includegraphics[width=8cm]{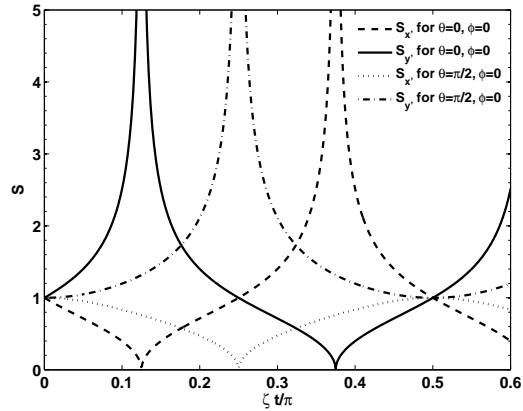}
         \caption{Squeezing S in the two-atom system is plotted versus $\eta=\zeta t$ in units of
                  $\pi$.
                  Note that for $\theta=\pi$, curves are same as for $\theta=0$ but the $S_{x^\prime}$
                  and $S_{y^\prime}$ are interchanged.}
        \end{center}
         \end{figure}

\section{Systems with $j>1$}
For a system having more than two two-level atoms, it is highly complicated to get an analytical
expression. However, the problem can be tackled somewhat numerically. \\

We are concerned about the evolution
\begin{equation}
\vert sss\rangle = S_{spin}\vert j,\xi\rangle
\end{equation}
where $j>1$. Quantum mechanical average of any spin operator or products of them, say $\hat O$,  is
given by
\begin{eqnarray}
\langle\hat O\rangle = \langle sss\vert\hat O\vert sss\rangle &=&
 \sum_{m^\prime=-j}^{j}\sum_{m^{\prime\prime}=-j}^j\langle sss\vert j,m^\prime\rangle\times \nonumber \\
                      & & \langle j,m^\prime\vert \hat O
\vert\j,m^{\prime\prime}\rangle\langle j,m^{\prime\prime}\vert sss\rangle .
\end{eqnarray}
The inner product $\langle j,m \vert sss\rangle$ is given by
\begin{eqnarray}
\langle j,m\vert sss\rangle &=& \langle j,m\vert S_{spin}\vert j,\xi\rangle   \nonumber \\
           & = & \sum_k \langle j,m\vert sss_k\rangle\langle sss_k\vert j,\xi\rangle e^{-i\lambda_k t}
\end{eqnarray}
where we have used $S_{spin}=\exp(-iHt)$ with $H$ being given by Eq. (12). The $\vert sss_k\rangle$ are the
eigenvectors of $H$ with eigenvalues $\lambda_k$.\\

The dynamics exhibit a rich variety of characteristics of spin
squeezing. In Fig. 2, we display $S_y^\prime$ for short interaction
time, that is, $\eta=\zeta t=\pi/20$. The striking feature there is
that, for $\theta=\pi/2$, $S_y^\prime$ oscillates just below the
line $S_y^\prime =1$ indicating spin squeezing for all values of
$j$. Thus the operator $S_{spin}$ is capable of squeezing a large
number of atoms if they are initially prepared in a coherent state
such that $\langle j, \xi\vert J_z \vert j,\xi\rangle=0$. The
distribution function for the initial condition $\theta=\pi/2$
takes the form
\begin{equation}
P(j,m)=\vert\langle j,m\vert j,\xi=1\rangle\vert^2 =
\frac{1}{2^{2j}}\frac{(2j)!}{(j+m)!(j-m)!}
\end{equation}
where we have taken $\phi =0$ for simplicity. It can be easily shown that
$P(j,m)$ peaks at $m=0$ by using the expression for digamma function, the derivative
of the factorial function. The state $\vert j,m=0\rangle$ has the property of
maximum correlation among indiviual spins  [19] which is exploited by the
operation $S_{spin}$ to squeeze out noise displayed in Fig. (2). This type
of behaviour has also been noticed in the spin squeezing properties of
pseuo-Hermitian operator discussed in Ref. [11].
This striking property of the spin squeezing operator persists for
longer interaction times too.

\vskip 0.4cm
\begin{figure}[h]
        \begin{center}
        \includegraphics[width=8cm]{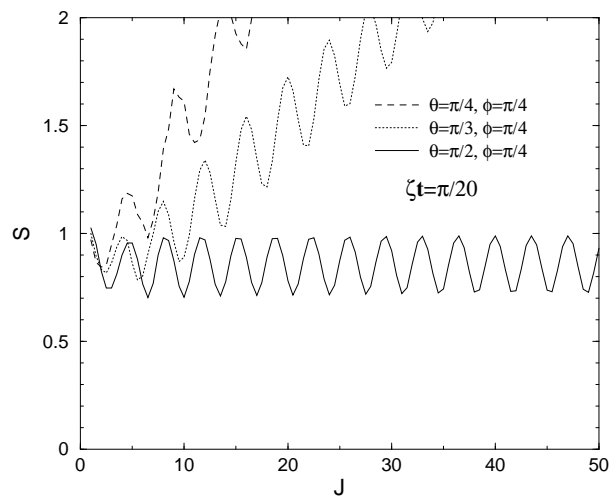}
         \caption{Squeezing in $y^\prime$ quadrature as a function of $j$.
                  Note that there is no squeezing in the other quadrature
                  for such interaction times.}
        \end{center}
         \end{figure}

\vskip 0.4cm
\begin{figure}[h]
        \begin{center}
        \includegraphics[width=8cm]{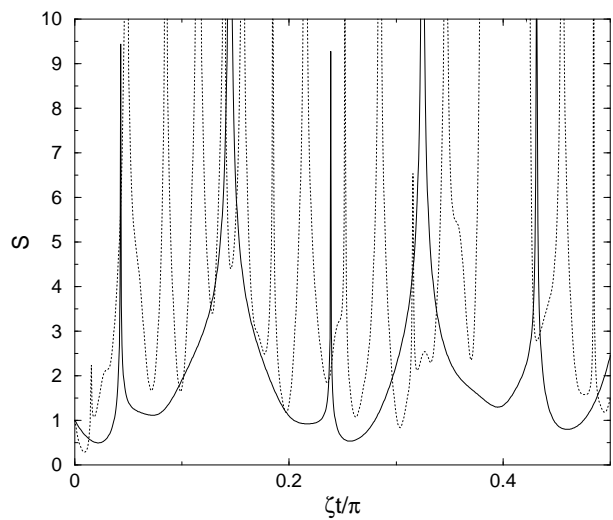}
         \caption{Oscillations in squeezing in $x^\prime$ quadrature as a function of $\eta=\zeta t$.
                  $\theta =\phi =0.$ The solid and dotted lines are for $j=4$ and $15$ respectively.}
        \end{center}
         \end{figure}

We display in Fig. 3 the variations in squeezing as a function of time for $j>1$. A
comparision with Fig. 1 indicates that the oscillations in squeezing with time
increase with $j$.

\section{Conclusions}

We have shown that operators in the form of exponentian of quadratic forms of spin
operators, $J_kJ_l$ have the spin squeezing properties when operated upon a spin
coherent state. We have shown in Fig. 2 that the operator can squeeze out noise from
an ensemble having a large number of spins initially prepared in a coherent state with
$\theta =\pi/2$ and $\phi =\pi/4$. As shown there, a deviation from this initial
condition reduces the number of squeezed spins drastically. These results, however,
would be influenced by the surrounding reservoirs which we plan to study
in a later publication.\\

The quadratic forms of spin operators have been discussed in the literature for quite
sometime. For example, a widely discussed interaction is the so-called Lipkin interaction
Hamiltonian [14]
\begin{equation}
H=G_1(J_{+}^2+J_{-}^2)+G_2(J_+J_- +J_-J_+).
\end{equation}
where $G_1$ and $G_2$ are the coupling constants representing the
two-body interactions. Indeed these two types of terms are the archetypes of the Hamiltonian
quadratic in the generators and related to what we expressed through Eqs. (11).
The operators in Eqs. (11) are special cases of the Lipkin
Hamiltonian. The $S_{spin}$ also appears in the Hamiltonian of a complex
magnetic molecule in a static magnetic field [20]. The operator $S^{\prime}_{spin}$ can be
realized in an optical system consisting of an ensemble of atoms in a high Q dispersive
cavity [13]. We have studied its behaviour in detail in Ref. [12]. Also, the exponent in
$S^{\prime}_{spin}$ describes atomic cooperation in a Dicke system [19] placed inside a
cavity and driven by an external laser field [21]. Hence, its Hamiltonian includes a term
which is proportional to $J_{z}^2$. Our recent study shows that this system exhibits spin
squeezing [22].\\


\end{document}